\documentclass[sigconf,screen]{acmart}
\usepackage[shortlabels]{enumitem}
\usepackage{listings, xcolor}
\renewcommand\footnotetextcopyrightpermission[1]{}
\usepackage{amsmath}
\usepackage{algorithm}
\usepackage{algpseudocode}
\usepackage{algorithmicx}
\usepackage{float}
\usepackage{pifont}
\usepackage{booktabs}
\usepackage{multirow}
\usepackage{bbding}
\usepackage[table]{xcolor}
\definecolor{lightblue}{rgb}{0.7, 0.85, 1}
\definecolor{verylightgray}{rgb}{.97,.97,.97}
\usepackage{tcolorbox}
\usepackage{minted}
\usepackage{tcolorbox}
\usepackage{xcolor}
\definecolor{vulred}{RGB}{150,60,60}
\definecolor{vulgreen}{RGB}{60,120,60}
\usepackage{enumitem}
\usepackage{listings}
\tcbuselibrary{skins, breakable}

\lstdefinelanguage{Solidity}{
  keywords={
    contract, function, modifier, event, mapping,
    address, uint, uint256, bool, public, private,
    external, internal, view, pure, payable,
    require, assert, if, else, for, while, return,
    emit, new
  },
  keywordstyle=\color{blue}\bfseries,
  ndkeywords={msg, sender, value, block, timestamp},
  ndkeywordstyle=\color{purple},
  identifierstyle=\color{black},
  sensitive=true,
  comment=[l]{//},
  morecomment=[s]{/*}{*/},
  commentstyle=\color{vulred},
  stringstyle=\color{orange},
  sensitive=true
}

\AtBeginDocument{%
  }

\begin{document}

\title{SCPatcher: Automated Smart Contract Code Repair via Retrieval-Augmented Generation and Knowledge Graph}


\author{Xiaoqi Li}
\email{csxqli@ieee.org}
\affiliation{%
  \institution{Hainan University}
  \city{Haikou}
  \country{China}
}

\author{Shipeng Ye}
\email{shipengye@hainanu.edu.cn}
\affiliation{%
  \institution{Hainan University}
  \city{Haikou}
  \country{China}
}

\author{Wenkai Li}
\email{cswkli@hainanu.edu.cn}
\affiliation{%
  \institution{Hainan University}
  \city{Haikou}
  \country{China}
}

\author{Zongwei Li}
\email{lizw1017@hainanu.edu.cn}
\affiliation{%
  \institution{Hainan University}
  \city{Haikou}
  \country{China}
}


\begin{abstract}
 Smart contract vulnerabilities can cause substantial financial losses due to the immutability of code after deployment. While existing tools detect vulnerabilities, they cannot effectively repair them. In this paper, we propose SCPatcher, a framework that combines retrieval-augmented generation with a knowledge graph for automated smart contract repair. We construct a knowledge graph from 5,000 verified Ethereum contracts, extracting function-level relationships to build a semantic network. This graph serves as an external knowledge base that enhances Large Language Model reasoning and enables precise vulnerability patching. We introduce a two-stage repair strategy, initial knowledge-guided repair followed by Chain-of-Thought reasoning for complex vulnerabilities. Evaluated on a diverse set of vulnerable contracts, SCPatcher achieves 81.5\% overall repair rate and 91.0\% compilation pass rate, substantially outperforming existing methods.
\end{abstract}



\maketitle

\section{Introduction}
Smart contracts are automated programs on blockchains \cite{zhang2025security} that execute when predetermined conditions are met \cite{li2025facial,bu2025smartbugbert}. As they manage significant financial assets, security is paramount. According to SlowMist's report~\cite{slowmist_hacked}, over \$2 billion has been lost due to the attacks on blockchain platforms in 2025. Developers typically rely on third-party auditing services to detect vulnerabilities before deployment, but effective repair remains challenging, especially for those lacking security expertise \cite{chaliasos2024smart}. In addition, manual patching is time-consuming, error-prone, and requires a deep understanding of both the vulnerability and the contract's business logic ~\cite{li2025beyond,zhu2024sybil}.

Existing detection tools (e.g., Oyente~\cite{luu2016making}, SmartBugs~\cite{di2023smartbugs}) identify vulnerabilities but provide limited repair guidance \cite{bobadilla2025automated}. Recent advances in Large Language Models (LLMs)~\cite{chen2025numscout,zhang2025acf,wang2024contracttinker,zhao2025recode,bouzenia2024repairagent} show promise for code generation and repair through their ability to understand programming patterns and semantic relationships. However, LLMs suffer from hallucinations, generating syntactically valid but semantically incorrect patches, particularly when handling domain-specific security patterns without external guidance.

\begin{figure*}[!t]
    \centering
    \includegraphics[width=0.95\linewidth]{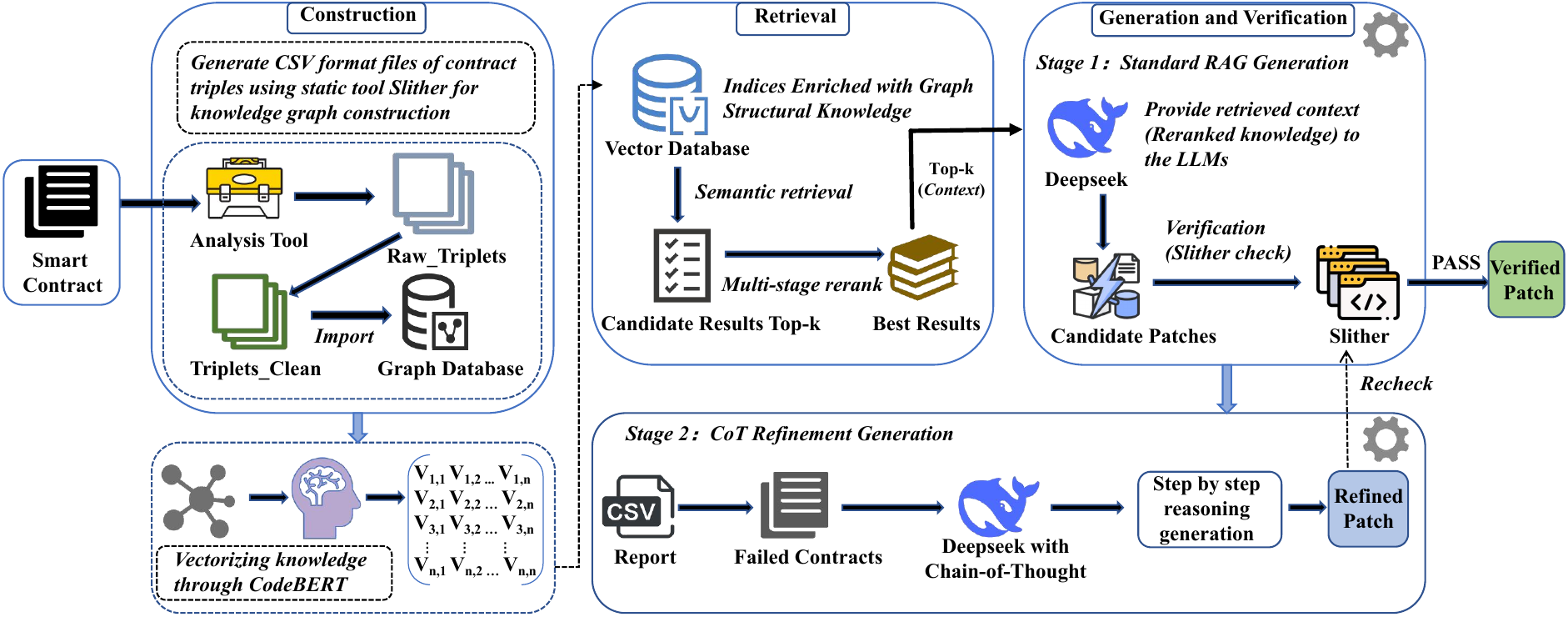}
    \caption{Overview of SCPatcher powered by Deepseek-V3. It retrieves knowledge information from a constructed external database, guiding Deepseek-V3 automatically to generate patches for vulnerable smart contracts.}
    \label{framework}
\end{figure*}

To address these limitations, we propose SCPatcher, which enhances LLM repair through structured knowledge retrieval. Our framework uses the static analysis tool Slither~\cite{feist2019slither} to extract contract relationships, builds a Knowledge Graph (KG) in Neo4j~\cite{zheng2025automating}, and embeds it using CodeBERT. This knowledge base guides LLMs to generate accurate repairs \cite{yuan2024designrepair}. For complex cases, we employ Chain-of-Thought (CoT)~\cite{wei2022chain} reasoning to handle deep logical dependencies, resolving contextual conflicts in vulnerability patching.\par

The main contributions of this paper are as follows:
\vspace{-3ex}

\begin{enumerate}
\item To the best of our knowledge, we provide the first systematic study of smart contract repair using RAG and KGs.\par
\item  We develop a hybrid retrieval mechanism integrating semantic search with trust-based reranking.\par
\item We propose a two-stage adaptive repair strategy combining knowledge-guided repair with CoT reasoning.\par
\item We conduct a series of experiments and evaluations demonstrating a superior performance over state-of-the-art.
\end{enumerate}

\vspace{-4.6ex}
\section{Background}
\subsection{Smart Contract Vulnerabilities}

Smart contracts written in Solidity are immutable once deployed \cite{wu2025exploring}. Vulnerabilities in these contracts can cause severe losses to blockchain systems. There are five kinds of common vulnerabilities in this study. \ding{172} \textbf{Integer Overflow/Underflow.} Arithmetic operations exceeding variable capacity. Earlier versions of Solidity (before 0.8.0) lacked built-in protections, allowing attackers to bypass authentication or disrupt token economics\cite{dinh2025enhancing}. \ding{173} \textbf{Reentrancy Attacks.} Malicious contracts exploit the gap between external calls and state updates to drain funds repeatedly. The DAO attack exemplifies this vulnerability\cite{dey2025reentrancy}. \ding{174} \textbf{Access Control Weaknesses.} Functions lacking proper restrictions allow unauthorized privilege escalation. Public initialization functions enable attackers to hijack administrator roles \cite{wang2022phyjacking}. \ding{175} \textbf{Timestamp Manipulation.} Miners can manipulate block timestamps to influence contract logic, particularly in random number generation for gaming contracts \cite{chen2025numscout}. \ding{176} \textbf{Unchecked Call Return Value.} External calls in Solidity return a boolean indicating success or failure. When contracts fail to check these return values, silent failures can occur, leading to incorrect state assumptions and potential fund loss ~\cite{yao2024compositional}. These vulnerabilities enable attackers to exploit contracts by manipulating their underlying logic. Consequently, effective auditing and repair are essential to ensure contract security prior to deployment\cite{zhang2025penetration}.

\subsection{LLMs for Code Repair}

In code security analysis, LLMs can parse function signatures, control flow, and data dependencies to detect vulnerabilities and generate repairs aligned with execution semantics\cite{annepaka2025large}.

However, LLMs face challenges including hallucinations (generating factually incorrect content), outdated knowledge requiring expensive retraining, and limited context windows. These limitations necessitate external knowledge augmentation.

\subsection{RAG with Knowledge Graphs}

Traditional RAG retrieves unstructured text, but knowledge graph-based RAG leverages semantic networks of entities and relationships \cite{peng2025graph}. This enables retrieval beyond surface similarity to capture logical connections. Knowledge graphs encode domain knowledge as nodes (entities) and edges (relationships), supporting contextual reasoning and dynamic updates. For code repair, this provides fact-checked, logically structured knowledge that reduces hallucination risks and enables more reliable vulnerability patching.

% However, these methods lack structured integration of domain knowledge and cannot capture semantic relationships in contracts.
% Without such structured knowledge, the critical contextual constraints and entity dependencies can be overlooked .

\section{Methodology}

Figure \ref{framework} shows SCPatcher's architecture, including database construction, hybrid retrieval, and generation with verification.\par

\subsection{Knowledge Graph Construction}

We collect 5,000 verified contracts from Etherscan\cite{etherscan2025} and use Slither \cite{feist2019slither} to extract entity-relationship triples (i.e., contracts, functions, variables, modifiers and their relationships like calls, owns, returns). These form a property graph stored in NetworkX~\cite{networkx2025}. We enrich the graph by identifying code clones, usage patterns, and control flow dependencies \cite{huang2023semantic}. CodeBERT embeds nodes into dense vectors to construct a vector database for efficient retrieval.

\subsection{Retrieval and Multi-Stage Reranking}

Given a vulnerable function, we encode it as query vector $v_{q}$ and perform a $k$-Nearest Neighbor (k-NN) search to retrieve a broad set (Top-50) of candidates $C_{init}$ based on semantic distance in Eq. \ref{eq:sem_dist}.

\begin{equation}
\label{eq:sem_dist}
S_{sem}(q, d) = \|\mathbf{v}_q - \mathbf{v}_d\|_2
\end{equation}

\begin{algorithm}[ht]
\caption{Multi-Stage Structure-Aware Reranking Algorithm}
\small
\label{alg:reranking}
\begin{algorithmic}[1]
\Require
    Initial candidate set $\mathcal{C}_{init}$;
    Query context $q$ with signature constraints $Sig_{req}$;
    Smoothing term $\epsilon$;
    Target context size $K$.
\Ensure
    Optimized context set $\mathcal{C}_{final}$.

\State \textbf{\textit{Stage 1: Syntactic Compatibility}}
\State $\mathcal{C}_{filt} \leftarrow \emptyset$
\For{$d \in \mathcal{C}_{init}$}
    \If{$Sig_{req} \subseteq d.metadata.signature$}
        \State $\mathcal{C}_{filt} \leftarrow \mathcal{C}_{filt} \cup \{d\}$
    \EndIf
\EndFor
\If{$\mathcal{C}_{filt} = \emptyset$} \Comment{Fallback mechanism to prevent empty set}
    \State $\mathcal{C}_{filt} \leftarrow \mathcal{C}_{init}$
\EndIf

\State \textbf{\textit{Stage 2: Trust-Enhanced Rescoring}}
\For{$d \in \mathcal{C}_{filt}$}
    \State $S_{sem} \leftarrow d.distance$ \Comment{Initial semantic distance}
    \State $f_{freq} \leftarrow d.metadata.\text{GUF}$
    \State $S_{final}(d) \leftarrow \frac{S_{sem}}{\ln(f_{freq} + \epsilon)}$ \Comment{Scoring function as defined in Eq. 2}
\EndFor
\State \textbf{Sort} $\mathcal{C}_{filt}$ by $S_{final}(d)$ in \textbf{ascending} order

\State \textbf{\textit{Stage 3: Diversity-Aware Deduplication}}
\State $\mathcal{C}_{final} \leftarrow \emptyset$
\State $\mathcal{S}_{seen} \leftarrow \emptyset$ \Comment{Set to track Semantic Clone Groups}
\For{$d \in \mathcal{C}_{filt}$}
    \State $id_{clone} \leftarrow d.metadata.\text{CloneID}$
    \If{$id_{clone} \notin \mathcal{S}_{seen}$ \textbf{or} $id_{clone} \text{ is Null}$}
        \State $\mathcal{C}_{final} \leftarrow \mathcal{C}_{final} \cup \{d\}$
        \State $\mathcal{S}_{seen} \leftarrow \mathcal{S}_{seen} \cup \{id_{clone}\}$
    \EndIf
    \If{$|\mathcal{C}_{final}| = K$}
        \State \textbf{break}
    \EndIf
\EndFor

\State \Return $\mathcal{C}_{final}$
\end{algorithmic}
\end{algorithm}

However, semantic similarity alone is insufficient for precise repair. We apply multi-stage reranking shown in the Algorithm \ref{alg:reranking}.

\begin{table*}[t]
\centering
\caption{Comparative Analysis of Repair Performance on 200 Smart Contract Vulnerabilities. We report the number of compiled/fixed contracts and their corresponding success rates. \textbf{ERR} denotes the repair success rate among compilable contracts, while \textbf{ORR} represents the overall success rate across the entire dataset.}

\label{tab:main_results}
\resizebox{\textwidth}{!}{% 自动缩放以适应页面宽度
\begin{tabular}{l|c|ccc|ccc|cc}
\toprule
\multirow{2}{*}{\textbf{Method}} & \multirow{2}{*}{\textbf{Paradigm}} & \multicolumn{3}{c|}{\textbf{Compilation Metrics}} & \multicolumn{3}{c|}{\textbf{Effective Repair Metrics}} & \multicolumn{2}{c}{\textbf{Overall Performance}} \\ 
\cmidrule(lr){3-5} \cmidrule(lr){6-8} \cmidrule(lr){9-10}
 &  & \textbf{$N_{comp}$} & \textbf{$N_{fail\_comp}$} & \textbf{CPR (\%)} & \textbf{$N_{fixed}$} & \textbf{$N_{fail\_fixed}$} & \textbf{ERR (\%)} & \textbf{Fixed / Total} & \textbf{ORR (\%)} \\ 
\midrule
\textbf{DirectFix \cite{napoli2023evaluating}} & Intuition & 153 & 47 & 76.5\%\small{\textbf{(\textcolor{teal}{$\uparrow$14.5\%})}} & 91 & 62 & 59.5\%\small{\textbf{(\textcolor{teal}{$\uparrow$30.1\%})}} & 91 / 200 & 45.5\%\small{\textbf{(\textcolor{teal}{$\uparrow$36.0\%})}} \\
\textbf{LogicRepair \cite{ai2025logicrepair}} & Reasoning & 176 & 24 & 88.0\%\small{\textbf{(\textcolor{teal}{$\uparrow$3.0\%})}} & 139 & 37 & 78.9\%\small{\textbf{(\textcolor{teal}{$\uparrow$10.7\%})}} & 139 / 200 & 69.5\%\small{\textbf{(\textcolor{teal}{$\uparrow$12.0\%})}} \\
\textbf{SelfRefine \cite{madaan2023self}} & Iteration & 132 & 68 & 66.0\%\small{\textbf{(\textcolor{teal}{$\uparrow$25.0\%})}} & 98 & 34 & 74.2\%\small{\textbf{(\textcolor{teal}{$\uparrow$15.4\%})}} & 98 / 200 & 49.0\%\small{\textbf{(\textcolor{teal}{$\uparrow$32.5\%})}} \\
\midrule
\textbf{SCPatcher (Ours)} & \textbf{Knowledge} & \textbf{182} & \textbf{18} & \textbf{91.0\%} & \textbf{163} & \textbf{19} & \textbf{89.6\%} & \textbf{163 / 200} & \textbf{81.5\%} \\ 
\bottomrule
\end{tabular}%
}

\end{table*}

\begin{itemize}
    \item\textbf{ Syntactic filtering.} Filtering out candidates that are incompatible with the target function signatures.
% Remove candidates incompatible with function signatures.

    \item\textbf{ Trust scoring.} Prioritizing the remaining candidates based on their Global Usage Frequency (GUF) in Eq. \ref{eq:ranking_score}.
% Prioritize frequently used patterns via Global Usage Frequency (GUF):

\begin{equation} 
\label{eq:ranking_score} 
S_{final}(d, q) = \frac{S_{sem}(\mathbf{v}_d, \mathbf{v}_q)}{\ln(f_{freq} + \epsilon)}
\end{equation}

    \item \textbf{ Diversity deduplication.} Select the top-$k$ candidates from distinct semantic clone groups.
\end{itemize}
This multi-stage reranking strategy ensures that the retrieved repair candidates are syntactically compatible with the target contract, validated through community usage, and semantically diverse.

\subsection{Generation and Verification}
\begin{figure}[ht]
    \centering
    \includegraphics[width=1\linewidth]{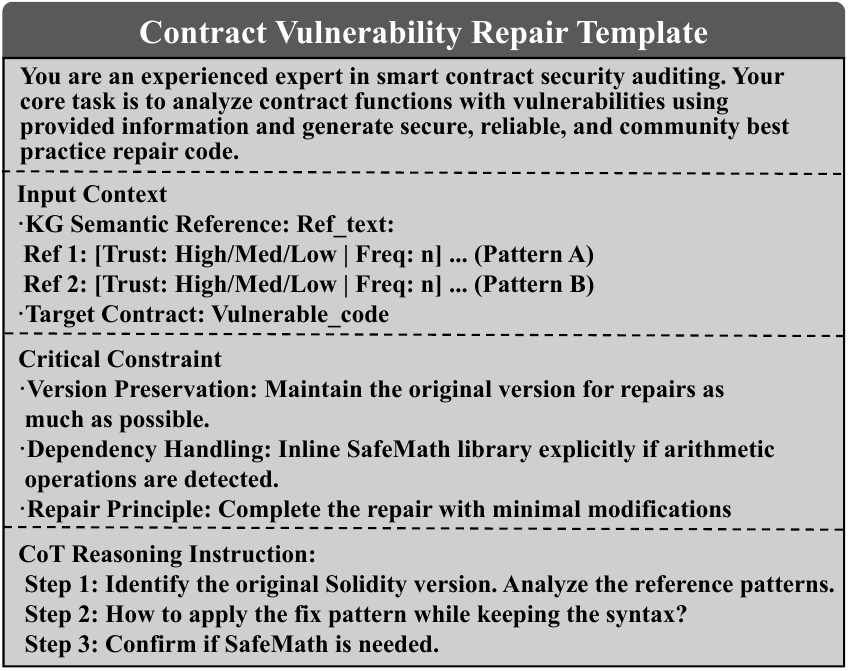}
    \caption{The Design of Prompt.}
    \vspace{-4ex}
    \label{prompt}
\end{figure}

We construct prompts, as shown in Figure \ref{prompt}, integrating retrieved knowledge with trust metrics and signature constraints. Deepseek V3 generates patches in two stages as follows.

\begin{itemize}
\item \textbf{Stage 1.} Knowledge-guided generation using retrieved references. Slither validates the output.

\item \textbf{Stage 2.} For failed cases, Chain-of-Thought reasoning guides step-by-step repair.

\end{itemize}

Both stages undergo Slither verification to ensure vulnerability elimination without introducing new issues.

\section{EXPERIMENT}

% \subsection{Experiment Setting}
All experiments were performed in a machine with NVIDIA GeForce RTX 4070Ti GPU (12GB) and Intel i7-12700 CPU.
We utilize 5000 verified contracts for knowledge graph construction. For evaluation, 200 vulnerable contracts from SmartBugs and open-source vulnerability repositories were tested, including reentrancy, integer overflow, access control, timestamp manipulation, and unchecked call return value. To prevent data leakage, we strictly deduplicated the 200 test contracts against the 5000 KG contracts using code hashing, ensuring no exact overlap.

\subsection{Performance Metrics}
To evaluate the performance of SCPatcher, we set up different measurement indicators, including $N_{comp}$ (Number of successful compilations), $N_{fail\_comp}$ (Number of compilation failures), $N_{fixed}$ (Number of successful repairs), and $N_{fail\_fixed}$ (Number of failed repairs). In the context of this study, a repair denotes vulnerability mitigation verified by static analysis, which does not guarantee full behavioral or functional correctness. Based on these counts, we calculate three key rates.

\begin{itemize}
\item \textbf{Compilation Pass Rate (CPR).} 

\begin{equation}
CPR = \frac{N_{comp}}{N} \times 100\%
\end{equation}

\item \textbf{Effective Repair Rate (ERR).} 

\begin{equation}
ERR = \frac{N_{fixed}}{N_{comp}} \times 100\%
\end{equation}

\item \textbf{Overall Repair Rate (ORR).} 

\begin{equation}
ORR = \frac{N_{fixed}}{N} \times 100\%
\end{equation}

\end{itemize}

\subsection{Baseline}
 To evaluate the performance of SCPatcher, we reproduce the prompting methods proposed by Napoli et al. \cite{napoli2023evaluating}, Ai et al. \cite{ai2025logicrepair}, and Madaan et al. \cite{madaan2023self}. In addition, for the convenience of statistics, we also name their methods DirectFix, LogicRepair, and SelfRefine respectively.\par

\subsection{Comparison Results}
As shown in the Table \ref{tab:main_results}, the experimental results demonstrate that our SCPatcher outperforms other baseline methods. SCPatcher achieves 81.5\% ORR, improving 12-36\% over baselines. High CPR of 91.0\% demonstrates syntactic correctness, while high ERR of 89.6\% confirms semantic accuracy.

\subsection{Ablation Study}
To evaluate the effectiveness of the key components in SCPatcher, we conduct a progressive ablation study. Table \ref{tab:ablation} shows that base LLM (Pure Generation) generates patches directly based on the vulnerable code without any external retrieval. Standard RAG represents the traditional retrieval-augmented generation approach. SCPatcher w/o CoT represents the first stage of repair that combines structural knowledge with reranking retrieval. SCPatcher (Full) is our proposed complete framework incorporating KG-enhanced retrieval, trust metrics, and the CoT refinement strategy. \par

\begin{table}[htbp]
\centering
% \vspace{-2ex}
\caption{Ablation Study Results. The performance of SCPatcher and each module.}
\vspace{-2ex}
\label{tab:ablation}
\resizebox{\columnwidth}{!}{%
\begin{tabular}{l|cc|ccc}
\hline
\textbf{Method} & \textbf{$N_{comp}$} & \textbf{$N_{fixed}$} & \textbf{CPR (\%)} & \textbf{ERR (\%)} & \textbf{ORR (\%)} \\ \hline
Base LLM (Pure)     & 141                 & 80                  & 70.5              & 56.7              & 40.0 \small{\textbf{(\textcolor{teal}{$\uparrow$41.5\%})}}       \\
Standard RAG        & 102                 & 56                  & 51.0              & 54.9              & 28.0 \small{\textbf{(\textcolor{teal}{$\uparrow$53.5\%})}}       \\
SCPatcher w/o CoT   & 180                 & 147                  & 90.0              & 81.7              & 73.5 \small{\textbf{(\textcolor{teal}{$\uparrow$8.0\%})}}  \\ \hline
\textbf{SCPatcher}  & \textbf{182}        & \textbf{163}         & \textbf{91.0}     & \textbf{89.6}     & \textbf{81.5}     \\ \hline
\end{tabular}%
}
\vspace{-2ex}
\end{table}

\subsection{Parameter Sensitivity Analysis}
\begin{figure}[h]
    \centering 
    % \vspace{-2ex}
        \centering
        \includegraphics[width=1\linewidth]{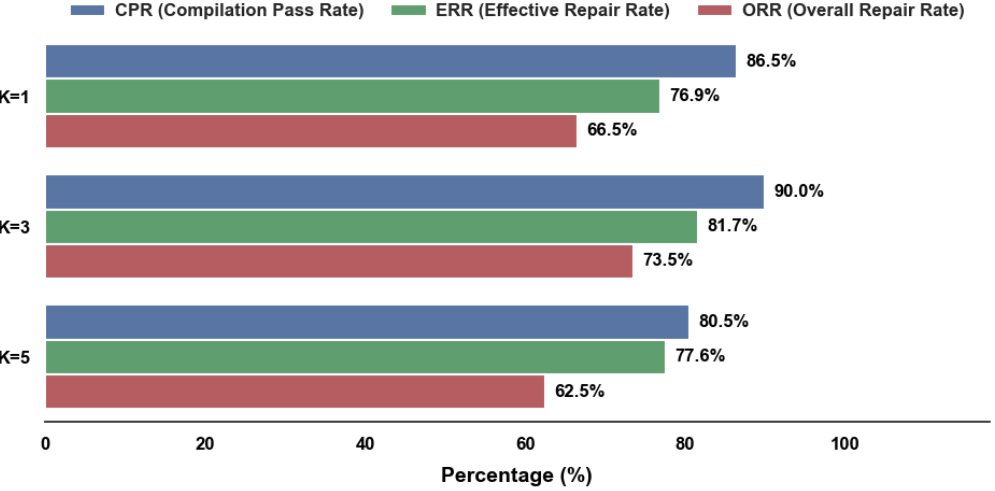}
        \vspace{-4ex}
        \caption{Sensitivity Analysis of the Number of Retrieved References ($K$) on Repair Performance.}
        \vspace{-2ex}
        \label{datasets}
    \vspace{-2ex}
\end{figure}

We evaluate SCPatcher's performance on the test set with $K \in \{1, 3, 5\}$ for the initial stage. As illustrated in Figure \ref{datasets}, SCPatcher achieves optimal intermediate performance at $K$=3, yielding repair patches with a CPR of 90.0\%, an ERR of 81.7\%, and an ORR of 73.5\% (prior to the final refinement stage).

\section{Discussion and Threats to Validity}

Through systematic experimentation, we identify three critical factors affecting repair performance as follows. \noindent\textbf{Retrieval approaches can effectively complement the reasoning-based methods.} LogicRepair leverages the CoT to achieve 69.5\% ORR, but knowledge-guided SCPatcher reaches 81.5\%, demonstrating external knowledge's necessity. \noindent\textbf{Contextual alignment significantly impacts repair quality.} In the ablation experiment of Table \ref{tab:ablation}, Standard RAG's poor performance of 28.0\% ORR shows that generic rules introduce noise, and domain-specific knowledge is essential. \noindent\textbf{Iterative refinement without proper guidance degrades patch correctness.} While iterative refinement is commonly assumed to improve output quality, our results challenge this assumption~\cite{liu2024deepseek,team2024gemini,lee2025evolving}. As shown in Table \ref{tab:main_results}, SelfRefine achieves only 49.0\% ORR, indicating that self-correction without external validation leads to overcorrection and degraded performance.

\textbf{Threats to Validity.} Relying on Slither risks patch-overfitting to static detectors without guaranteeing behavioral correctness. While effective for pattern-based bugs, our function-level graph and static checks may struggle with complex, non-pattern vulnerabilities like cross-contract dependencies or deep business logic errors.

% It is generally believed that reflection and improvement can enhance quality. However, SelfRefine's performance (49.0\% ORR) reveals that self-criticism without external validation causes overcorrection.

% In our experiments, we also identify limitations of SCPatcher. Its effectiveness in patching vulnerabilities is highly dependent on the coverage of the knowledge base. For vulnerabilities absent from the knowledge base or newly discovered, RAG retrieval performance may decline, potentially leading LLMs to generate erroneous code. 

Based on these limitations, we plan to improve SCPatcher in three aspects. First, we will enrich the knowledge graph by integrating a broader range of vulnerability corpora and real-world repair patterns to improve retrieval precision. Second, we aim to integrate dynamic execution and test cases into the verification process to rigorously ensure semantic correctness. Finally, we will explore adaptive repair strategies that dynamically adjust the generation process based on compiler feedback and execution results.

\vspace{-2ex}
\section{Related Work}

In the domain of traditional program repair, several studies have been conducted~\cite{jiang2021cure,li2022dear,yang2024cref,bouzenia2024repairagent,zhao2025recode}. Early approaches like CURE \cite{jiang2021cure} used Neural Machine Translation (NMT) with Abstract Syntax Trees. DEAR \cite{li2022dear} combined fault localization with deep learning. For recent methods integrating LLMs, CREF \cite{yang2024cref} used conversational interaction, RepairAgent \cite{bouzenia2024repairagent} employed autonomous planning, and ReCode \cite{zhao2025recode} applied retrieval-augmented generation. Regarding the smart contract repair, ContractWard \cite{wang2020contractward} pioneered machine learning for vulnerability detection. SCRepair \cite{yu2020smart} introduced gas-aware repair optimization. sGuard+ \cite{gao2024sguard+} combined machine learning detection with defensive patterns. 
Recent advances in LLM-based repair tools include ContractTinker \cite{wang2024contracttinker}, which employs task decomposition strategies, and ACFix \cite{zhang2025acf}, which specializes in addressing access control vulnerabilities. These techniques have laid the foundation for emerging smart contract repair approaches. In particular, the advancement of LLMs has accelerated the development of automated repair by enabling more sophisticated code understanding and generation capabilities. Different from these works, we construct the structured knowledge that captures the semantic relationships in contracts, which is critical to integrate the contextual constraints and entity dependencies for repairing~\cite{he2025ontological}. 

\section{Conclusion}
% This paper proposed SCPatcher, a LLM-based framework for automated smart contract repair. By constructing a domain knowledge graph from contract triples and employing hybrid retrieval with trust-based reranking, SCPatcher provides LLMs with precise, validated repair guidance. A two-stage approach combining knowledge-guided generation with Chain-of-Thought reasoning addresses both standard and complex vulnerabilities. Experimental results demonstrate that SCPatcher outperforms baselines in automatic repair. Our study not only validates the effectiveness of the framework but also provides a new idea for leveraging LLMs with structured domain knowledge for smart contract repair.
This paper presents SCPatcher, an LLM-based framework for automated smart contract repair. SCPatcher constructs a domain-specific knowledge graph from contract triples and employs a hybrid retrieval mechanism with trust-based reranking to provide LLMs with precise, validated repair guidance. The framework adopts a two-stage repair process that combines knowledge-guided patch generation with Chain-of-Thought reasoning, enabling effective handling of both common and complex vulnerabilities. Experimental evaluation demonstrates that SCPatcher significantly outperforms existing baseline approaches across multiple repair metrics. Beyond validating the proposed framework, this work establishes a promising paradigm for integrating structured domain knowledge with LLMs to advance automated program repair in the smart contract domain.

\bibliographystyle{ACM-Reference-Format}
\bibliography{main}

@article{li2025facial,
  title={Facial Recognition Leveraging Generative Adversarial Networks},
  author={Li, Zhongwen and Li, Zongwei and Li, Xiaoqi},
  journal={arXiv preprint arXiv:2505.11884},
  year={2025}
}

@misc{slowmist_hacked,
  title        = {{SlowMist Hacked: Blockchain Hack Archives}},
  author       = {{SlowMist}},
  year         = {2025},
  howpublished = {\url{https://hacked.slowmist.io/}},
  note         = {Accessed: 2025-10-23}
}

@article{bobadilla2025automated,
  title={Do automated fixes truly mitigate smart contract exploits?},
  author={Bobadilla, Sofia and Jin, Monica and Monperrus, Martin},
  journal={IEEE Transactions on Software Engineering},
  year={2025},
  publisher={IEEE}
}

@article{wu2025exploring,
  title={Exploring vulnerabilities and concerns in solana smart contracts},
  author={Wu, Xiangfan and Xing, Ju and Li, Xiaoqi},
  journal={arXiv preprint arXiv:2504.07419},
  year={2025}
}

@inproceedings{dey2025reentrancy,
  title={Reentrancy Vulnerability in the Blockchain Ecosystem: Historical Attacks, Detection Approaches, and Mitigation Strategies},
  author={Dey, Tamoghna and Lenka, Rakesh Kumar and Senapati, Shruti and Mishra, Debani Prasad and Mallick, Soubhagya Ranjan},
  booktitle={2025 International Conference on Networks and Cryptology (NETCRYPT)},
  pages={1307--1311},
  year={2025},
  organization={IEEE}
}

@article{dinh2025enhancing,
  title={Enhancing Smart Contract Security Through DevSecOps: An Adaptive Approach for Vulnerability Detection},
  author={Dinh, Nghia and Hoang, Vinh Truong and Van, Bay Nguyen and Huong, Thien Ho and Hong, Ha Duong Thi and Trung, Hau Nguyen and Trung, Kiet Tran},
  journal={IEEE Access},
  year={2025},
  publisher={IEEE}
}

@article{zhang2025security,
  title={Security analysis of ponzi schemes in ethereum smart contracts},
  author={Zhang, Chunyi and Wei, Qinghong and Li, Xiaoqi},
  journal={arXiv preprint arXiv:2510.03819},
  year={2025}
}

@inproceedings{wang2022phyjacking,
  title={PHYjacking: Physical Input Hijacking for Zero-Permission Authorization Attacks on Android.},
  author={Wang, Xianbo and Shi, Shangcheng and Chen, Yikang and Lau, Wing Cheong},
  booktitle={Proceedings of the Network and Distributed System Security Symposium (NDSS)},
  year={2022}
}

@article{chen2025numscout,
  title={NumScout: Unveiling Numerical Defects in Smart Contracts using LLM-Pruning Symbolic Execution},
  author={Chen, Jiachi and Shao, Zhenzhe and Yang, Shuo and Shen, Yiming and Wang, Yanlin and Chen, Ting and Shan, Zhenyu and Zheng, Zibin},
  journal={IEEE Transactions on Software Engineering},
  year={2025},
  publisher={IEEE}
}

@book{yao2024compositional,
  title={Compositional Security for Smart Contracts},
  author={Yao, Siqiu},
  year={2024},
  publisher={Cornell University}
}

@article{annepaka2025large,
  title={Large language models: a survey of their development, capabilities, and applications},
  author={Annepaka, Yadagiri and Pakray, Partha},
  journal={Knowledge and Information Systems},
  volume={67},
  number={3},
  pages={2967--3022},
  year={2025},
  publisher={Springer}
}

@article{peng2025graph,
  title={Graph retrieval-augmented generation: A survey},
  author={Peng, Boci and Zhu, Yun and Liu, Yongchao and Bo, Xiaohe and Shi, Haizhou and Hong, Chuntao and Zhang, Yan and Tang, Siliang},
  journal={ACM Transactions on Information Systems},
  volume={44},
  number={2},
  pages={1--52},
  year={2025},
  publisher={ACM New York, NY}
}

@inproceedings{jiang2021cure,
  title={Cure: Code-aware neural machine translation for automatic program repair},
  author={Jiang, Nan and Lutellier, Thibaud and Tan, Lin},
  booktitle={Proceedings of the IEEE/ACM 43rd International Conference on Software Engineering (ICSE)},
  pages={1161--1173},
  year={2021},
  organization={IEEE}
}

@inproceedings{li2022dear,
  title={Dear: A novel deep learning-based approach for automated program repair},
  author={Li, Yi and Wang, Shaohua and Nguyen, Tien N},
  booktitle={Proceedings of the 44th international conference on software engineering},
  pages={511--523},
  year={2022}
}

@inproceedings{yang2024cref,
  title={Cref: An llm-based conversational software repair framework for programming tutors},
  author={Yang, Boyang and Tian, Haoye and Pian, Weiguo and Yu, Haoran and Wang, Haitao and Klein, Jacques and Bissyand{\'e}, Tegawend{\'e} F and Jin, Shunfu},
  booktitle={Proceedings of the 33rd ACM SIGSOFT International Symposium on Software Testing and Analysis (ISSTA)},
  pages={882--894},
  year={2024}
}

@article{bouzenia2024repairagent,
  title={Repairagent: An autonomous, llm-based agent for program repair},
  author={Bouzenia, Islem and Devanbu, Premkumar and Pradel, Michael},
  journal={arXiv preprint arXiv:2403.17134},
  year={2024}
}

@inproceedings{zhao2025recode,
  title={ReCode: Improving LLM-based Code Repair with Fine-Grained Retrieval-Augmented Generation},
  author={Zhao, Yicong and Chen, Shisong and Zhang, Jiacheng and Li, Zhixu},
  booktitle={Proceedings of the 34th ACM International Conference on Information and Knowledge Management (CIKM)},
  pages={4368--4378},
  year={2025}
}

@article{wang2020contractward,
  title={Contractward: Automated vulnerability detection models for ethereum smart contracts},
  author={Wang, Wei and Song, Jingjing and Xu, Guangquan and Li, Yidong and Wang, Hao and Su, Chunhua},
  journal={IEEE Transactions on Network Science and Engineering},
  volume={8},
  number={2},
  pages={1133--1144},
  year={2020},
  publisher={IEEE}
}

@article{yu2020smart,
  title={Smart contract repair},
  author={Yu, Xiao Liang and Al-Bataineh, Omar and Lo, David and Roychoudhury, Abhik},
  journal={ACM Transactions on Software Engineering and Methodology (TOSEM)},
  volume={29},
  number={4},
  pages={1--32},
  year={2020},
  publisher={ACM New York, NY, USA}
}

@article{gao2024sguard+,
  title={sGuard+: Machine learning guided rule-based automated vulnerability repair on smart contracts},
  author={Gao, Cuifeng and Yang, Wenzhang and Ye, Jiaming and Xue, Yinxing and Sun, Jun},
  journal={ACM Transactions on Software Engineering and Methodology},
  volume={33},
  number={5},
  pages={1--55},
  year={2024},
  publisher={ACM New York, NY}
}

@inproceedings{wang2024contracttinker,
  title={Contracttinker: Llm-empowered vulnerability repair for real-world smart contracts},
  author={Wang, Che and Zhang, Jiashuo and Gao, Jianbo and Xia, Libin and Guan, Zhi and Chen, Zhong},
  booktitle={Proceedings of the 39th IEEE/ACM International Conference on Automated Software Engineering (ASE)},
  pages={2350--2353},
  year={2024}
}

@article{zhang2025acf,
  title={ACFix: Guiding LLMs with Mined Common RBAC Practices for Context-Aware Repair of Access Control Vulnerabilities in Smart Contracts},
  author={Zhang, Lyuye and Li, Kaixuan and Sun, Kairan and Wu, Daoyuan and Liu, Ye and Tian, Haoye and Liu, Yang},
  journal={IEEE Transactions on Software Engineering},
  year={2025},
  publisher={IEEE}
}

@article{huang2023semantic,
  title={Semantic-enriched code knowledge graph to reveal unknowns in smart contract code reuse},
  author={Huang, Qing and Liao, Dianshu and Xing, Zhenchang and Zuo, Zhengkang and Wang, Changjing and Xia, Xin},
  journal={ACM Transactions on Software Engineering and Methodology},
  volume={32},
  number={6},
  pages={1--37},
  year={2023},
  publisher={ACM New York, NY}
}

@inproceedings{napoli2023evaluating,
  title={Evaluating chatgpt for smart contracts vulnerability correction},
  author={Napoli, Emanuele Antonio and Gatteschi, Valentina},
  booktitle={2023 IEEE 47th Annual Computers, Software, and Applications Conference (COMPSAC)},
  pages={1828--1833},
  year={2023},
  organization={IEEE}
}

@inproceedings{ai2025logicrepair,
  title={LogicRepair: An Empirical Study on Automated Repair of Smart Contract Logic Vulnerabilities Based on Large Language Models},
  author={Ai, Mingchao and Wang, Kai and Wang, He and Zheng, Chen and Zhang, Yuqing},
  booktitle={Proceedings of the 11th IEEE International Conference on Privacy Computing and Data Security (PCDS)},
  pages={261--266},
  year={2025},
  organization={IEEE}
}

@article{madaan2023self,
  title={Self-refine: Iterative refinement with self-feedback, 2023},
  author={Madaan, Aman and Tandon, Niket and Gupta, Prakhar and Hallinan, Skyler and Gao, Luyu and Wiegreffe, Sarah and Alon, Uri and Dziri, Nouha and Prabhumoye, Shrimai and Yang, Yiming and others},
  journal={URL https://arxiv. org/abs/2303.17651},
  year={2023}
}

@article{li2025beyond,
  title={Beyond the Hype: A Large-Scale Empirical Analysis of On-Chain Transactions in NFT Scams},
  author={Li, Wenkai and Li, Zongwei and Li, Xiaoqi and Zhang, Chunyi and Zhang, Xiaoyan and Zhang, Yuqing},
  journal={arXiv preprint arXiv:2512.01577},
  year={2025}
}

@article{bu2025smartbugbert,
  title={Smartbugbert: Bert-enhanced vulnerability detection for smart contract bytecode},
  author={Bu, Jiuyang and Li, Wenkai and Li, Zongwei and Zhang, Zeng and Li, Xiaoqi},
  journal={arXiv preprint arXiv:2504.05002},
  year={2025}
}

@inproceedings{chaliasos2024smart,
  title={Smart contract and defi security tools: Do they meet the needs of practitioners?},
  author={Chaliasos, Stefanos and Charalambous, Marcos Antonios and Zhou, Liyi and Galanopoulou, Rafaila and Gervais, Arthur and Mitropoulos, Dimitris and Livshits, Benjamin},
  booktitle={Proceedings of the 46th IEEE/ACM International Conference on Software Engineering (ICSE)},
  pages={1--13},
  year={2024}
}

@article{yuan2024designrepair,
  title={Designrepair: Dual-stream design guideline-aware frontend repair with large language models},
  author={Yuan, Mingyue and Chen, Jieshan and Xing, Zhenchang and Quigley, Aaron and Luo, Yuyu and Luo, Tianqi and Mohammadi, Gelareh and Lu, Qinghua and Zhu, Liming},
  journal={arXiv preprint arXiv:2411.01606},
  year={2024}
}

@article{zhang2025penetration,
  title={Penetration testing for system security: Methods and practical approaches},
  author={Zhang, Wei and Xing, Ju and Li, Xiaoqi},
  journal={arXiv preprint arXiv:2505.19174},
  year={2025}
}

@article{he2025ontological,
  title={An Ontological Knowledge-Driven Smart Contract Framework for Implicit Bridge Preservation Decision Making},
  author={He, Chuanni and Liu, Min and Hsiang, Simon M and Pierce, Nicholas and Megahed, Samuel and Godfrey, Asa},
  journal={Journal of Construction Engineering and Management},
  volume={151},
  number={4},
  pages={04025008},
  year={2025},
  publisher={American Society of Civil Engineers}
}

@inproceedings{feist2019slither,
  title={Slither: a static analysis framework for smart contracts},
  author={Feist, Josselin and Grieco, Gustavo and Groce, Alex},
  booktitle={Proceedings of the IEEE/ACM 2nd International Workshop on Emerging Trends in Software Engineering for Blockchain (WETSEB)},
  pages={8--15},
  year={2019},
  organization={IEEE}
}

@inproceedings{luu2016making,
  title={Making smart contracts smarter},
  author={Luu, Loi and Chu, Duc-Hiep and Olickel, Hrishi and Saxena, Prateek and Hobor, Aquinas},
  booktitle={Proceedings of the ACM SIGSAC conference on computer and communications security (CCS)},
  pages={254--269},
  year={2016}
}

@inproceedings{di2023smartbugs,
  title={Smartbugs 2.0: An execution framework for weakness detection in ethereum smart contracts},
  author={Di Angelo, Monika and Durieux, Thomas and Ferreira, Jo{\~a}o F and Salzer, Gernot},
  booktitle={Proceedings of the 38th IEEE/ACM International Conference on Automated Software Engineering (ASE)},
  pages={2102--2105},
  year={2023},
  organization={IEEE}
}

@misc{etherscan2025,
      author = "Etherscan",
      title = "The Ethereum Browser",
      year = "2026",
      url = "https://etherscan.io/",
}

@misc{networkx2025,
      author = "NetworkX",
      title = "Network Analysis in Python",
      year = "2026",
      url = "https://networkx.org/en/",
}

@article{liu2024deepseek,
  title={Deepseek-v3 technical report},
  author={Liu, Aixin and Feng, Bei and Xue, Bing and Wang, Bingxuan and Wu, Bochao and Lu, Chengda and Zhao, Chenggang and Deng, Chengqi and Zhang, Chenyu and Ruan, Chong and others},
  journal={arXiv preprint arXiv:2412.19437},
  year={2024}
}

@article{team2024gemini,
  title={Gemini 1.5: Unlocking multimodal understanding across millions of tokens of context},
  author={Team, Gemini and Georgiev, Petko and Lei, Ving Ian and Burnell, Ryan and Bai, Libin and Gulati, Anmol and Tanzer, Garrett and Vincent, Damien and Pan, Zhufeng and Wang, Shibo and others},
  journal={arXiv preprint arXiv:2403.05530},
  year={2024}
}

@article{lee2025evolving,
  title={Evolving deeper llm thinking},
  author={Lee, Kuang-Huei and Fischer, Ian and Wu, Yueh-Hua and Marwood, Dave and Baluja, Shumeet and Schuurmans, Dale and Chen, Xinyun},
  journal={arXiv preprint arXiv:2501.09891},
  year={2025}
}

@article{wei2022chain,
  title={Chain-of-thought prompting elicits reasoning in large language models},
  author={Wei, Jason and Wang, Xuezhi and Schuurmans, Dale and Bosma, Maarten and Xia, Fei and Chi, Ed and Le, Quoc V and Zhou, Denny and others},
  journal={Advances in neural information processing systems},
  volume={35},
  pages={24824--24837},
  year={2022}
}

@article{zheng2025automating,
  title={Automating construction contract review using knowledge graph-enhanced large language models},
  author={Zheng, Chunmo and Wong, Saika and Su, Xing and Tang, Yinqiu and Nawaz, Ahsan and Kassem, Mohamad},
  journal={Automation in Construction},
  volume={175},
  pages={106179},
  year={2025},
  publisher={Elsevier}
}

@article{zhu2024sybil,
  title={Sybil attacks detection and traceability mechanism based on beacon packets in connected automobile vehicles},
  author={Zhu, Yaling and Zeng, Jia and Weng, Fangchen and Han, Dan and Yang, Yiyu and Li, Xiaoqi and Zhang, Yuqing},
  journal={Sensors},
  volume={24},
  number={7},
  pages={2153},
  year={2024},
  publisher={MDPI}
}

\appendix

\end{document}